# RESCUE: Interdependent Challenges of Reliability, Security and Quality in Nanoelectronic Systems

M. Jenihhin[1], S. Hamdioui[2], M. Sonza Reorda[3], M. Krstic[4], P. Langendörfer[4], C. Sauer[5], A. Klotz[5], M. Huebner[6], J. Nolte[6], H. T.Vierhaus[6], G. Selimis[7], D. Alexandrescu[8], M. Taouil[2], G. J. Schrijen[7], J. Raik[1], L. Sterpone[3], G. Squillero[3], Z. Dyka[4]

[1]Tallinn University of Technology, Estonia; [2]Delft University of Technology, The Netherlands; [3]Politecnico di Torino, Italy;
[4]IHP – Leibniz-Institut für innovative Mikroelektronik, Germany; [5]Cadence Design Systems GmbH, Germany;
[6]BTU Cottbus-Senftenberg, Germany; [7]Intrinsic ID B.V., The Netherlands; [8]IROC Technologies, France
maksim.jenihhin@taltech.ee

*Abstract*— The recent trends for nanoelectronic computing systems include machine-to-machine communication in the era of Internet-of-Things (IoT) and autonomous systems, complex safety-critical applications, extreme miniaturization of implementation technologies and intensive interaction with the physical world. These set tough requirements on mutually dependent extra-functional design aspects. The H2020 MSCA ITN project RESCUE is focused on key challenges for reliability, security and quality, as well as related electronic design automation tools and methodologies. The objectives include both research advancements and cross-sectoral training of a new generation of interdisciplinary researchers. Notable interdisciplinary collaborative research results for the first half-period include novel approaches for test generation, soft-error and transient faults vulnerability analysis, cross-layer fault-tolerance and error-resilience, functional safety validation, reliability assessment and run-time management, HW security enhancement and initial implementation of these into holistic EDA tools.

## I. INTRODUCTION

The long-lasting and steady technology scaling has enabled significant advances in functionality density and architectural solutions. These, in turn, lead to diverse application fields for integrated circuits from security RF-ID chips and biomedical implanted nanoelectronic devices to many-core processors for artificial intelligence, autonomous driving and cloud servers with billions of transistors integrated. Nanoelectronic systems, containing both hardware and embedded software components, are being combined today into the Internet of Things and Cyber-Physical Systems (CPSs) and, ultimately, represent the physical backbone of our increasingly digitized world.

More and more nanoelectronic systems are being deployed in life-critical application domains, such as healthcare, transportation, automotive and security, serving societal needs. Here, the impact and consequences of in-field failures, security attacks or hardware bugs and defects can be catastrophic. Reliability, quality and security cannot be treated anymore as standalone aspects and also have inherent tradeoffs with a set of application constraints, cost-efficiency, energy consumption, performance of the system and its safety requirements [2], [3], [35]. Due to today's market driven applications and demands, the requirements are becoming a necessity even for consumer electronics such as smart phones and wearables. To underpin the next generation implementation technologies and rescue the steady growth of nanoelectronic systems' functionality, new methodologies and electronic design automation (EDA) tools for interdisciplinary and multi-scale design, modelling, and analysis are urgently needed.

As an example, the current technology used in the high-end smartphones include a processor with over 8.5 billion transistors. The 7nm TSMC FinFET technology of the A13 Bionic, used to develop one of the most complex embedded Systems-on-Chip (SoCs) in the mobile communication domain of these days, includes a multi-core system and accelerators for artificial intelligence. Such hardware architecture is approaching the physical limits of technology and EDA scalability, highly vulnerable to faults and, therefore, needs specific mechanisms defined at design time to operate the SoC reliably, safely and securely. Today, these mechanisms go way beyond just a set of redundant gates or data paths. The online fault detection and repair has to be implemented very carefully in order to keep the performance and power requirements of the chip. Recent methods for this purpose are already a must in the state-of-the-art chipsets. The next generation of chips will need even more specific and efficient realizations to build dependable and resilient hardware.

The H2020 Marie Skłodowska-Curie Innovative Training Network action RESCUE [1] establishes a network for an interdisciplinary and cross-sectoral research and training [22] for future European engineers and researchers. Traditionally, the research and training in Europe for these highly interdependent challenges in nanoelectronic system design is fragmented and performed by scattered communities. The cross-sectoral consortium of RESCUE is well-balanced in terms of academic and industrial research facilities to tackle the reliability, security and quality challenges in a holistic manner. The industrial sector behind this initiative includes innovative and award-winning European SMEs from the areas of nano-electronics reliability and security - IROC Technologies and Intrinsic ID. The large companies on board are Cadence Design Systems, a global leader in electronic design automation, and Robert Bosch, the European automotive electronics flagship. The latter supports the ETN as a partner organization. As cutting-edge research institution, Leibniz-Institute IHP serves as a bridge for knowledge transfer between the sectors. The academic sector is represented by Delft, Brandenburg and Tallinn Universities of Technology, and Politecnico di Torino. RESCUE was launched on April 1, 2017 and will last for 4 years with the total budget 3.76 MEUR, as a contribution by the European Commission.

The rest of this paper is organized as follows: Section II outlines the objectives and concepts of the project, Section III

This research was supported in part by project RESCUE funded from the EU H2020 programme under the MSC grant agreement No.722325 and by EU through the European Structural and Regional Development Funds.

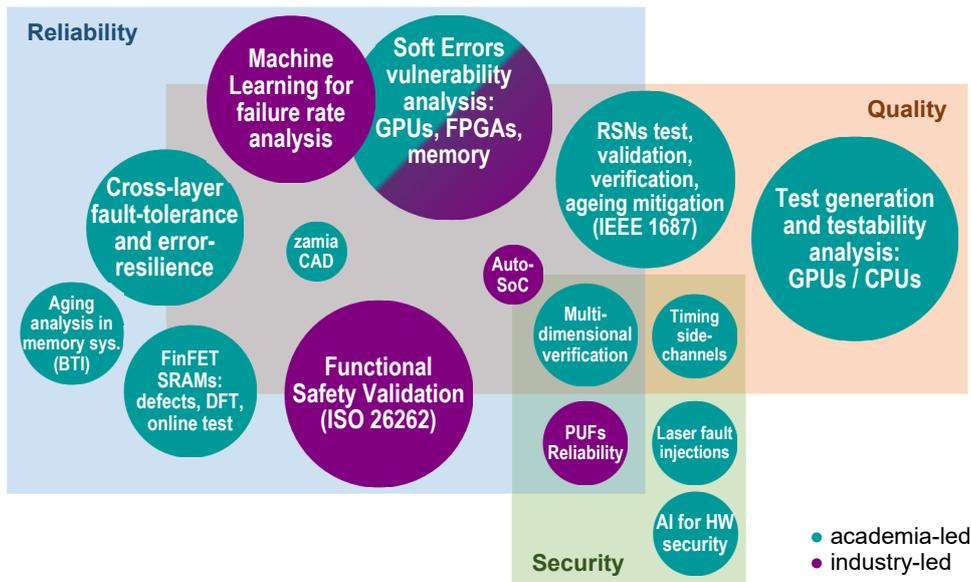

Fig 1. Distribution of the RESCUE project's interdisciplinary collaborative research results for the first half-period

outlines the main interdisciplinary research results of the project and Section IV discusses the corresponding experimental framework. Finally, Section VI draws the conclusions.

## II. OBJECTIVES AND CONCEPTS OF THE PROJECT

The research objective of the project is to address major technological and scientific challenges in an interdisciplinary area involving quality, reliability, security and EDA tools, which will enable and enhance the design and manufacturability of complex systems at smaller technology nodes. Here, within the scope of the project these terms are fixed to the following. *Reliability of nanoelectronic systems* is subject to threats during the system's lifetime in the field such as wear-out or ageing defects and errors coming from the environment, e.g. radiation-caused soft errors. *Quality of nanoelectronic systems* can be compromised by threats at time zero of the system's life. These include design errors, manufacturing defects, nanometer-technology process variation, etc. and are addressed by means of pre- and post-silicon functional validation, test and diagnosis. *Security of nanoelectronic systems* can be compromised by attacks on design IP (intellectual property), data asset and design functionality. The key actions here are secure design of hardware and embedded software parts and accurate security evaluation. *EDA tools and methodologies* for secure, correct and reliable nanoelectronics are developed with a holistic approach.

As the second objective, the project provides early-stage researchers (ESRs) with intensive cross-sectoral training in the involved disciplines [22]. This is supported by dedicated events [8] and benefits from a portfolio of technical and transferable skills courses available in the network (e.g. [9]). The project is implemented by defining 15 detailed ESRs' individual research projects while keeping both objectives under consideration.

## III. INTERDISCIPLINARY RESEARCH

A distribution of the RESCUE project's interdisciplinary collaborative research results for the first half-period is illustrated in Fig. 1. The size of the "bubbles" is proportional to the number of publications and preliminary results. The main accent in the first half-period was made on individual techniques e.g. for the reliability, quality and fault-tolerance aspects of electronic systems. Several interdisciplinary initiatives addressing the security aspect are work-in-progress with publishable results expected soon.

### A. Test generation and testability analysis

General Purpose Graphics Processing Units (GPGPUs) have been considered in the frame of RESCUE because they represent an interesting case. Originally, they were introduced for applications, such as graphics and gaming, where reliability is a minor concern. In the following years, they started to be adopted for High Performance Computing (HPC) and, more recently, for safety-critical applications in the automotive domain. On the other side, they normally exploit advanced semiconductor technologies, which are known to be more prone to faults and to critical effects, such as aging. Hence, solutions to make them reliable enough for these applications are urgently needed.

The research work done in RESCUE concerning GPGPUs focused, first, on developing solutions able to effectively detect possible permanent faults arising during the operational life [11], [41], [42]. The proposed techniques belong to the general category of functional ones (Software-based Self-test) and were evaluated resorting to an existing GPGPU model (FlexGrip) which has been significantly improved and expanded in the frame of the project [43]. Thanks to the availability of the improved FlexGrip model, for the first time in the literature we have been able to quantitatively assess the effectiveness of a test solution on a GPGPU. Secondly, a joint work focused on the automatic identification of the functionally untestable faults inside a GPGPU [46]. This step is crucial to correctly estimate the fault coverage achieved by any test method, and allows to reduce the cost for functional fault simulation. A similar activity was also performed targeting more conventional RISC processors [23], [28], [33]. Finally, RESCUE researchers analyzed the impact of permanent and transient faults when some typical applications running on GPGPUs are considered [25], also evaluating the impact on reliability and performance stemming from different software encoding styles [40].

*B. Soft- Error and Transient Faults vulnerability analysis*

Among other reliability threats, transient faults, such as Single-Event Upsets (SEUs) in sequential/state logic and Single-Event Transients (SETs) in combinatorial logic, are known to contribute significantly to the overall failure rate of the system, possibly exceeding the set reliability targets. As an example, standard flip-flops and SRAM memories, manufactured in relatively recent technologies (down to the latest CMOS bulk processes) exhibit error rates of hundreds of FITs (events per a billion working hours per megabit). Complex circuits using such cells can easily overshoot the 10 FIT target mandated by the ISO 26262 for an automotive ASIL D application.

RESCUE proposes methods [12], [13], [14] to study circuits' sensitivity to transient faults and single events mainly caused by radiation particle induced faults occurring at the circuit layout level and impacting the circuit's cell behavior and the subsequent propagation into the system, provoking observable failures at the system level.

Characterization of radiation-induced soft errors and the reliability analysis on advanced integrated circuits such as GPGPUs is a key pillar to individuate new techniques for the modeling of SEUs and SETs on circuit. A suitable model to cope with the investigation in harsh environments has been presented in [43]. On the other side, due to the technology scaling, lower supply voltages and higher operational frequencies, SETs are becoming a big concern also to specific resources such as the clock distribution network (CDN) or the reset circuitry. [54] proposes a comprehensive framework for the analysis of the impact of CDN SETs to the functional behavior of a circuit.

The reliability assessment process is usually accomplished with different types of fault injection methods like exhaustive and random. The first one is obviously ultimate in terms of accuracy but very cumbersome in terms of resources, time, EDA licenses and so on, making this approach unfeasible on medium and large circuits. The random fault injection method provides a solution to avoid unreasonable costs while allowing for accuracy (or statistical significance) on the proposed scope using mathematical and statistical methods. Along with these, RESCUE researchers also explore the use of Machine Learning techniques [31], [55], [56], [57], [57] for reliability and functional safety evaluation, allowing fast and accurate fault, error and failure metric extraction and evaluation.

*C. Cross-layer fault tolerance and error resilience*

Concerning fault tolerance in current and future hardware, our research focuses on fault detection and complementary fault repair mechanisms. Fault handling at lower levels close to the area where the error occurred allows to avoid high, often unacceptable, latencies implied if decisions are made by a higher-level component on chip. On the other hand, modules for error detection and correction often can do more complex analysis and even track a "history" of faults and methods to repair them. Methods using Artificial Intelligence [4] can be also envisioned and are under our investigation. A higher-level component receiving information from low-level monitoring elements is able to decide on a more abstract level the behavior of the chip and informs the real-time Operating System about the status of the underlying hardware. In RESCUE, we develop a "meet in the middle" approach [52] where low-level monitoring and correction is accomplished with a high-level fault management. This addresses two major goals, i.e. a low-latency reaction to faults and a more complex and flexible fault management, considering in future also AI algorithms.

The important aspect of cross-layer fault tolerance is effective sensing and decision making about the potential system reconfiguration based on the actual environmental and intrinsic changes. In RESCUE, we are working on the development of novel sensing mechanisms for radiation monitoring. The specific characteristic of this approach is lies in the usage of the available memory resources on the chip, that are functionally utilized also for the SEU monitoring [38], [39], [53]. These monitors could be integrated with the other monitor types, i.e. fault monitors, ageing (such as Bias Temperature Instability - BTI or Hot Carrier Injection – HCI phenomena), temperature sensors, and used for intelligent system management.

*D. Functional safety validation*

The increasing usage of electronic systems in the automotive domain and their growing complexity due to applications such as autonomous driving causes a shift in the traditional design flows and pushing compliance to standards such as ISO26262 down to the semiconductor chain. With this, functional safety needs to become a first-class citizen throughout the full design flow. This concerns not only the safe function of the system in the field, but also the design and software tools involved into its development. Our proposed vendor-independent methodology helps improving the confidence in fault analysis tools by combining the strengths of Automatic Test Pattern generators (ATPGs), Formal methods and Fault Injection (FI) simulation to automatically verify tools and detect any errors in their fault classification [20], [48], [50].

Critical in the design process is the efficient evaluation of the design's robustness in coping with random hardware failures, including all its aspects in digital, analog, or software domains. In early stages of the flow, techniques for supporting architects and reliability experts in performing FMECA (Failure Mode, Effects and Criticality Analysis) are introduced, as well as for formally proving that certain critical states are never reached [19]. In later stages of the flow, assumptions and estimations about the systems function in the presence of faults need to be verified, requiring fault injection campaigns. Depending on underlying fault models and on the design characteristics these campaigns typically run 100s of 1000s of simulations at gate level in the digital domain. With millions of design components susceptible for random faults and elaborated verification environments, this requires significant efforts and time. Our work on dynamic slicing aims at pruned fault lists and smarter injection to save some of these efforts [49], [51]. How to extend FuSa (Functional Safety) verification in terms of its fault models as well as into the analog domain are also active areas of research in the RESCUE project.

*E. Reliability assessment and run-time management*

An interesting example showing how much the different aspects are correlated in current design flows is represented by Reconfigurable Scan Networks (RSNs), such as those supported by the IEEE1149 and IEEE1687 standards. These circuit structures are introduced to ease and optimize the access to internal registers used to calibrate, debug, and test the circuit. Hence, they have an extra-functional purpose. However, they may also be prone to design errors and manufacturing faults. For this reason, RESCUE early-stage researchers are working towards the development of effective solutions to test [15], [16],

[17], [30], [44], validate [29], [47] and diagnose faults [45] in RSNs, which are suitable to be integrated into future EDA tools. Moreover, since they allow accessing the inside of circuits, they must be protected against unwanted accesses, thus raising concerns about their security. In RESCUE, we have also studied the impact of BTI aging [36] on these critical infrastructures often used to organize access to embedded instruments and system health management.

Typically, to address the time-dependent degradation (aging), dedicated hardware mitigation schemes are applied. In another direction of our current work, we are also using existing on-chip resources to mitigate the memory BTI-induced aging, as the dominant phenomenon for the current technologies. As the baseline, we rely on our previous results demonstrated that by running programs on the processor design that the unbalanced logic paths can be rejuvenated using software [7]. In the context of this project, we extend this approach by using the processor to mitigate (parts of) the memory system. The idea is to embed additional instructions to the program to ensure a balanced stress of different parts of the memory. Our preliminary results show that the address decoder can be mitigated very well [24].

As SRAM memory dominates the chip area it is critical to ensure that this functions properly throughout its lifetime. To increase the fault coverage, we are looking at new approaches to model the new defects of the Fin Field-Effect Transistor (FinFET) technology and analyze their impact on both quality and reliability. To realize this, we are working on a methodology based on Technology Computer-Aided Design (TCAD) that can accurately capture the behavior of unique FinFET manufacturing defects in FinFET SRAMs. Each defect is modelled by altering the physical structure of FinFET devices to include unwanted characteristics, such as cracks on the channel or bended fins. These devices are then simulated for electrical analysis and their behavior on the cells are observed. This characterized behavior is used in the project for development of novel specialized Design for Testability (DfT) and mitigation schemes for FinFET memories [26]. In addition to that, we are also working on efficient (online) test solutions. To monitor the health status of an SRAM, we investigated efficient ways to monitor the status of cells using on-chip current sensors [10], [27]. The idea is to compare the response of different cells with each other and from there identify defective or weak cells. This allows for testing all defects simultaneously while using a limited number of operations only.

*F. Hardware security analysis and enhancement*

Due to the nature of applications such as critical infrastructure and the Internet of Things, side channel analysis attacks are becoming a serious threat. This is due to the fact that devices are deployed in the field without any protection means, i.e. they can be stolen and attacked using side channel analysis attacks in a lab. Side channel attacks (SCAs) take advantage from the fact that the behavior of crypto implementations can be observed and provides hints that simplify revealing keys. So new means to prevent or at least to increase the effort to run successful SCAs are needed.

A specific type of SCA are fault injection attacks. With fault injection, the attacker's objective is to change a critical value or to change the flow of a program. In order to cope with those attacks the behavior of the devices under attack needs to be understood. This is the reason why we are investigating [18] physical laser-based fault injection attacks in the IHP technologies available in the RESCUE network. For test structures we could show that fault injections switching a single transistor at least in the 250nm technology are successful and repeatable. This means changing states of identified registers that allow/prevent access to sensitive data such as keys can be changed by an attacker. New elements are under development that are aimed to prevent such attacks from being successful. Experiments are planned for the near future. Moreover, in RESCUE we follow an AI-based strategy against fault injection attacks. We are developing a new strategy based on neural networks which can detect faults in the program flow of critical functions such as the crypto engines. The neural network is trained with non-faulty traces only and hence has the potential to not only detect existing fault attacks but also future attacks.

Apart the active SCA fault injection attacks, we investigate the passive SCAs as well. In these attacks, the attacker passively listens in one of the side channels (e.g. time or power) in the hope that some sensitive data leaks. We have developed a verification framework for timing SCA. SCA data leakage has been identified using the framework and countermeasures have been taken [34]. At the moment, we validate our framework by investigating attacks on further hardware designs and there is a work in progress to introduce extra side channel attacks (e.g. power) to the framework.

Finally, we are investigating secure and low-cost ways of storing keys. In modern systems, the use of non-volatile memories for key storage gives room for attacks, since keys are always available in memory. One of the solutions to tackle this issue is Physical Unclonable Functions (PUFs) [6]. With PUFs the random uncontrollable manufacturing parameters of the device can be used to create a unique identifier and a cryptographic key root. However, due to technology scaling, there is a need to validate PUF designs under these emerging technologies. Over the last years, new technologies with different parameters and structures are proposed and manufactured, such as FinFET. We have developed a simulation framework and an analytical mathematical model for FinFET SRAM PUFs in order to be able to investigate reliability and entropy performance. First results are expected to be published soon.

IV. EXPERIMENTAL FRAMEWORK

*A. Holistic EDA Framework*

One of the goals of the RESCUE project is to establish holistic EDA methodologies along with corresponding tool flows for the interdependent design aspects of reliability, security and quality [32] (see Fig. 2). To understand the interference of functional and extra-functional design aspects the project has performed a comprehensive study of the state of the art [35], [21]. The cutting-edge academic research ideas are planned to be first implemented into experimental frameworks and have a potential to be integrated into standard industrial tool design flows from Cadence or reliability and functional safety-oriented EDA tools from IROC. zamiaCAD [5] is one of the academic open-source experimental platforms supported by several early-stage researchers. In practice, EDA toolsets and methodologies can be application specific targeting at systems' domains such as autonomous systems [37] (including the automotive domain and robotics), space applications, IoT edge devices, security-enabling HW, fault management infrastructures (IJTAG/RSNs), specific architectures (NoCs, many-cores, HMPSoCs), etc.

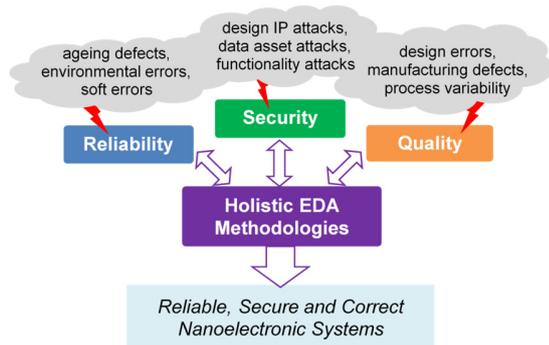

Fig. 2. RESCUE holistic approach to EDA tools and methodologies

Several key components are needed for cross-layer design methodologies that address multi-level circuit design flows. Firstly, extra-functional information, such as technology fault data, environment-induced events rates, etc., must be generated, consumed and exchanged transparently and safely. The project uses and significantly extends the *Reliability Information Interchange Format (RIIF)* to support the new design paradigms. Secondly, designing and testing many ideas and principles can benefit from "big data" information such as fault injection information and circuit reliability data. However, this is not always easy to obtain or generate. As a work in progress, RESCUE aims at generating and providing to the community large databases with the results of fault simulation campaigns and reliability analysis of complex circuits that can help further cross-layer design techniques. Lastly, the adoption of community-driven open-source formats, tools and methodologies is a fundamental principle of the project.

*B. Open-Source Automotive Benchmark Auto-SoC*

The development of Autonomous Vehicles applications, where a system failure could cause life-threatening situations, entails state-of-the-art challenges on different aspects of system development. Concerns with reliability, security, quality, and compliance to safety standards are of high priority. This scenario requires the adoption of new techniques and methodologies that will facilitate the development and verification of these applications. Different organizations are working to close the technological gap for Autonomous Vehicles. However, in order to assess the quality of the proposed solutions, it is necessary to compare the results against what is applied in the industry. Nowadays, development life-cycles and verification techniques applied by industry are not disclosed, and each big player in the automotive sector has its own methodologies and tools. In addition, automotive hardware and software solutions are seldomly available. This is a challenge for researchers that are not able to verify their work on representative designs or to quantitatively assess the quality of their results in a comparable manner. For that reason, there is a high demand for a suite of open-source benchmarks that would enable research on the different aspects of Automotive applications development.

For such a benchmark suite to be considered a valid solution, it should be characterized as:

- *Representative*: Based on the requirements of real-world systems;
- *Comparable*: Must allow comparability between different proposed methodologies and results;
- *Open*: All the components should be open-source allowing the exploration of bottlenecks;
- *Modular*: Consent for future growth and modification on components.

To gather the requirements for representative Automotive SoCs benchmarks, in frames of the RESCUE project we have analyzed several commercial solutions, in cooperation with some major players in the area. The analysis considered the main characteristics of commercial automotive SoCs, to identify key aspects that are common to all, and therefore, should be implemented in a possible new benchmark suite. The evaluation was focused on: (1) Architecture: common characteristics (e.g. CPUs, memories, automotive protocols); (2) Safety: what components of the SoCs are considered for functional safety compliance and what safety mechanisms are implemented; (3) Security: what security features are available; (4) Other: common available peripherals (e.g. communication networks, GPU, Audio/Video DSPs).

Based on this evaluation, the basic functionalities and architecture were defined for a benchmark suite named Automotive SoC (*AutoSoC*), corresponding to a SoC hardware based on the OR1200 CPU and including application-specific, memory and peripheral blocks. The RT-level synthesizable Verilog model of the hardware is available in a number of configurations, including different safety mechanisms to increase reliability, such as LockStep for the CPU and ECCs for the memories and a security block. Remarkably, the suite also includes some software to be run on the benchmark hardware, including a Linux Operating System version (with drivers for the peripherals) as well as a few representative applications.

*C. Chip Demonstrator*

Not all novel approaches for design reliability and quality assessment and enhancement can be accurately evaluated using simulation-in-software. As an early work in progress, joint activities of the project partners are pursuing also the design of a common silicon demonstrator. The demonstrator shall include the reliability, security and quality aware hardware and software IPs from the consortium, but also the contribution in terms of design flow improvements, as well as test approach enhancements. It may take advantage of the IHP facilities and can provide involved ESRs with practical experience of a real nanoelectronic system implementation flow.

V. CONCLUSIONS

H2020 MSCA ITN project RESCUE is focused on key challenges for reliability, security and quality, as well as related electronic design automation tools and methodologies. The first collaborative research results include a set of very promising approaches. The next step is to integrate these into a holistic EDA tools flow, open-source benchmark suits and a physical chip demonstrator.